\documentclass[aps,prb,twocolumn,showpacs,floatfix,superscriptaddress,10pt]{revtex4-2}
\usepackage{amsmath,amsxtra,amssymb,latexsym,amscd,amsthm}
\usepackage{indentfirst}
\usepackage{float}
\usepackage{epsfig}
\usepackage[mathscr]{eucal}
\usepackage{graphicx}
\usepackage{color}
\usepackage{mathtools}
\usepackage{xcolor}
\usepackage{titlesec}
\usepackage{natbib}
\usepackage[margin=1in]{geometry}
\usepackage{hyperref} % Load hyperref as the last package
\usepackage{bigints}
\usepackage{amsfonts}
\usepackage{boites,boites_exemples}
\usepackage[framemethod=TikZ]{mdframed}
\usepackage{amssymb}
\usepackage{stmaryrd}
\usepackage[utf8]{inputenc}
\usepackage[T1]{fontenc}
\usepackage[thinc]{esdiff}
\newcommand{\lt}{\left(}  
\newcommand{\rt}{\right)} 
\newmdenv[linecolor=black,skipabove=\topsep,skipbelow=\topsep,
leftmargin=-10pt,rightmargin=-10pt,
innerleftmargin=10pt,innerrightmargin=10pt]{mybox}

\begin{document}
\title{Spheroid rolling up on diverging inclines}

\author{Khanh P. M. Hoang}
\affiliation{Newton Grammar School, Hanoi 10000, Vietnam}
\email{hoangphamminhkhanh@gmail.com}
\author{Duy V. Nguyen}
\affiliation{Phenikaa School of Computing, Phenikaa University, Hanoi 12116, Vietnam}
\affiliation{Phenikaa Institute for Advanced Study, Phenikaa University, Hanoi 12116, Vietnam}
\email{duy.nguyenvan@phenikaa-uni.edu.vn}

\begin{abstract}
Objects rolling upward on diverging inclined rails exhibit a counterintuitive behavior that has intrigued physicists and students alike. Among the most well-known is the double-cone paradox, where a cone appears to roll uphill due to the geometry of the rails. In this paper, we extend this idea to more general rigid bodies, including spheres and ellipsoids. We examine the physical conditions under which this motion occurs, and we derive the necessary geometric and energetic constraints using analytical mechanics. We then validate our theoretical models through carefully designed experiments.
\end{abstract}
\maketitle
\section{Introduction}
In classical mechanics, few demonstrations have garnered as much educational interest as the so-called double-cone paradox—an experiment in which a double cone, when placed on a set of diverging inclined rails, appears to roll "uphill." While visually perplexing, this apparent defiance of gravity is, in fact, a product of the system’s geometry. As the rails diverge, the contact points between the cone and rails move closer to the cone’s axis of symmetry. This results in the center of mass of the cone descending, despite the cone’s ends appearing to gain elevation. The motion thus obeys the principle of conservation of energy and the natural tendency of systems to minimize potential energy. Historically, this paradox has been used in classroom settings to challenge students’ preconceptions about rolling motion and gravitational force \cite{gandhi2005} \cite{deluca2011}.\\
The origins of the double-cone experiment trace back to the Renaissance, with possible conceptual contributions by Leonardo da Vinci, and it has remained a topic of pedagogical fascination for centuries \cite{suciu2016}. In recent decades, variants of this demonstration have been introduced using spheres, ellipsoids, and even mechanical constructs such as geared double-cones. Among these is Gardner’s variation, which replaces the double cone with a ping-pong ball rolling between two pencils arranged in a V-shape. This adaptation preserves the essential geometric mechanism and introduces a simpler object—the sphere—making the paradox even more accessible while still retaining its core instructional value \cite{deluca2011}.\\
Despite the extensive use of these paradoxes in educational demonstrations, there has been relatively limited quantitative exploration of how this behavior generalizes to different shapes such as spheres and ellipsoids. The sphere, in particular, introduces a symmetry that simplifies certain aspects of analysis, yet also removes the cone's intrinsic radius-gradient that drives the changing elevation of contact points. This raises compelling theoretical questions: Under what geometric and energetic conditions can a sphere exhibit motion analogous to that of the double cone? How does the mass distribution and curvature of an ellipsoid influence its trajectory on diverging rails? What are the mathematical constraints on the rail geometry that lead to apparent uphill motion?\\
These questions serve as the central motivation for the present study. We extend the classical analysis of the double-cone system to include general rigid bodies—specifically, spheres and ellipsoids—rolling without slipping on a pair of rails that both incline and diverge. By constructing a full three-dimensional parametric model of the rails and incorporating geometric constraints into the body’s rolling motion, we derive the conditions under which the center of mass lowers, even as the object appears to ascend. Using Lagrangian mechanics and analytical expressions for potential energy, we identify key parameters—such as rail inclination ($\theta$), divergence angle ($\alpha$), and the aspect ratio of the body—that dictate the direction and stability of the motion.\\
The theoretical framework is further supported by experimental validation. Using precisely constructed setups with diverging rails and a range of test objects—including spheres and ellipsoids of various radii and mass distributions—we examine the equilibrium points, stability conditions, and range of motion of each body type. These experiments not only verify the analytical predictions but also reveal the presence of unstable equilibrium points, a hallmark of the underlying mechanical paradox \cite{deluca2011}.\\
Beyond its theoretical significance, this research holds value in education and engineering. For physics instruction, the generalization from cone to sphere or ellipsoid offers an opportunity to teach mechanics through a unified framework grounded in geometric reasoning. For engineering, such systems exemplify passive mechanical design principles, where geometry alone governs motion—principles that may be harnessed in robotics, motion-guided transport, or even energy harvesting mechanisms such as wave-powered generators using double-cone systems \cite{suciu2018}.\\
By rigorously analyzing the dynamics of a sphere and ellipsoid rolling on diverging inclined guides, this paper builds a bridge between an enduring classroom demonstration and a broader class of physical systems governed by geometry. In doing so, it reveals not only the mathematical elegance but also the practical utility of seemingly paradoxical motion.
\section{Unstable Equilibrium Position}
\subsection{The famous double-cone problem}
We will now propose a solution to the famous double-cone ascending problem. The two parameters of the double-cone are radius $R$ and height $h$.\\
Let the center of mass of the double-cone be denoted by $\mathrm{C}$, with coordinates $(x_{\mathrm{C}}, 0, z_{\mathrm{C}})$. Let the contact point on the rail be $\mathrm{M}$, with coordinates $(x_\mathrm{M}, y_\mathrm{M}, z_\mathrm{M})$. Again, the rail is described by the parametric equation: \[x=t, y=t \tan \dfrac{\alpha}{2}, z=t \tan \theta.\]
Since $\mathrm{M}$ lies on the rail, its coordinates must satisfy this relation. Therefore, we obtain the following constraints of $x_\mathrm{M}$, $y_\mathrm{M}$, and $z_\mathrm{M}$:
\begin{align*}
    y_\mathrm{M} &= x_\mathrm{M} \tan \dfrac{\alpha}{2}, \\
    z_\mathrm{M} &= x_\mathrm{M} \tan \theta.
\end{align*}
\definecolor{xfqqff}{rgb}{0.4980392156862745,0,1}
\definecolor{qqqqff}{rgb}{0,0,1}
\begin{tikzpicture}[scale=1.2]
\draw [line width=1pt] (1.6195410477840575,1.1947528636972198) circle (0.9660802535568241cm);
\draw [shift={(0,0)},line width=1pt,fill=black,fill opacity=0.1] (0,0) -- (0:0.6035119314829717) arc (0:21.80140948635181:0.6035119314829717) -- cycle;
\draw [line width=1pt,color=qqqqff] (0,0)-- (5,2);
\draw [line width=0.6pt][->] (0,0)-- (5.58,0);
\draw [line width=0.6pt][->] (0,0)-- (0,2.55);
\draw [line width=0.6pt,dash pattern=on 1pt off 1pt] (5,2)-- (5,0);
\draw (-0.2,0) node[anchor=north west] {$\text{O}$};
\draw (5.5,0.2) node[anchor=north west] {$x$};
\draw (-0.15,2.8) node[anchor=north west] {$z$};
\draw (0.6,0.35) node[anchor=north west] {$\theta$};
\draw (1.4,1.6) node[anchor=north west] {$\text{C}$};
\draw (1.1,1) node[anchor=north west] {$\text{M}$};
\begin{scriptsize}
\draw [fill=black] (1.6195410477840575,1.1947528636972198) circle (1.5pt);
\draw [fill=black] (1.3962476033010771,0.5584990413204309) circle (1.5pt);
\end{scriptsize}
\end{tikzpicture}
\begin{figure}
    \centering
    \begin{tikzpicture}[scale=0.5]
\draw [shift={(-6,0)},line width=0.6pt,fill=black,fill opacity=0.1] (0,0) -- (3.3664606634298013:2.317040343178322) arc (3.3664606634298013:22.380135051959574:2.317040343178322) -- cycle;
\draw [shift={(-6,0)},line width=1pt,fill=black,fill opacity=0.1] (0,0) -- (15.255118703057775:1.2743721887480772) arc (15.255118703057775:33.690067525979785:1.2743721887480772) -- cycle;
\draw [line width=0.6pt] (-6,0)-- (5,0);
\draw [line width=0.6pt] (5,0)-- (5,3);
\draw [line width=0.6pt] (0,4)-- (5,3);
\draw [line width=0.6pt] (-6,0)-- (2.5,3.5);
\draw [line width=0.6pt,dash pattern=on 1pt off 1pt] (0,1)-- (5,0);
\draw [line width=0.6pt,dash pattern=on 1pt off 1pt] (-6,0)-- (0,1);
\draw [line width=0.6pt,dash pattern=on 1pt off 1pt] (-6,0)-- (2.5,0.5);
\draw [line width=0.6pt,dash pattern=on 1pt off 1pt] (2.5,0.5)-- (2.5,3.5);
\draw [line width=0.6pt] (0,4)-- (0,1.6363636363636365);
\draw [line width=0.6pt,dash pattern=on 1pt off 1pt] (0,1.6363636363636365)-- (0,1);
\draw [line width=0.6pt][->] (2.5,0.5)-- (7.0753383673176735,0.7691375510186866);
\draw [line width=0.6pt][<-] (-6.001047982108175,4.403095046430594)-- (-6,0);
\draw [line width=0.6pt][->] (-6,0)-- (-6.93952270504735,2.7731126329046543);
\draw [line width=1pt,color=qqqqff] (-6,0)-- (-3.6056611402638623,1.5962259064907585);
\draw [line width=1pt,dash pattern=on 1pt off 1pt,color=qqqqff] (-3.6056611402638623,1.5962259064907585)-- (-2.1927037778936533,2.538197481404231);
\draw [line width=1pt,color=qqqqff] (-2.1927037778936533,2.538197481404231)-- (0,4);
\draw [line width=1pt,color=qqqqff] (-6,0)-- (-3.0885277935960995,0.794037874473791);
\draw [line width=1pt,dash pattern=on 1pt off 1pt,color=qqqqff] (-3.0885277935960995,0.794037874473791)-- (-1.357901375917633,1.2660268974770092);
\draw [line width=1pt,color=xfqqff] (-1.357901375917633,1.2660268974770092)-- (5,3);
\draw (-6.15,1.3) node[anchor=north west] {$2\phi$};
\draw (7,1.2) node[anchor=north west] {$x$};
\draw (-6.5,0) node[anchor=north west] {$\mathrm{O}$};
\draw (-7.5,3.5) node[anchor=north west] {$y$};
\draw (-6.4,5.060328056148228) node[anchor=north west] {$z$};
\draw (-3.75,0.9) node[anchor=north west] {$\theta$};
\draw (-2.5180772328281558,2.187198030607096) node[anchor=north west] {$\mathrm{C}$};
\draw (-3.5,3) node[anchor=north west] {$\mathrm{M}$};
\draw [rotate around={22.6552327086798:(-2.379287372214417,1.5452164665798716)},line width=1pt,fill=black,fill opacity=0.39] (-2.379287372214417,1.5452164665798716) ellipse (1.0353585288986062cm and 0.5154175745066195cm);
\draw [line width=1pt] (-3.311933840525477,1.1035721434902994)-- (-3.697941529620088,3.7934137185501955);
\draw [line width=1pt] (-3.697941529620088,3.7934137185501955)-- (-1.4277749653640712,1.9530136722828928);
\draw [line width=1pt] (-1.4277749653640712,1.9530136722828928)-- (-1.3488636317280194,-0.6741577436985262);
\draw [line width=1pt] (-1.3488636317280194,-0.6741577436985262)-- (-3.311933840525477,1.1035721434902994);
\draw [line width=1pt] (-1.4277749653640712,1.9530136722828928)-- (-1.3488636317280194,-0.6741577436985262);
\begin{scriptsize}
\draw [fill=black] (-2.3937070024423277,1.484944175464924) circle (1.5pt);
\draw [fill=black] (-2.9614285545204293,2.0257142969863806) circle (1.5pt);
\draw [fill=black] (-2.4655620999103594,0.9639376091153564) circle (1.5pt);
\end{scriptsize}
\end{tikzpicture}

\definecolor{qqqqff}{rgb}{0,0,1}
\begin{tikzpicture}[scale=1.2]
\draw [shift={(0,0)},line width=0.6pt,fill=black,fill opacity=0.1] (0,0) -- (-14.036243467926479:0.5840960489399741) arc (-14.036243467926479:14.036243467926479:0.5840960489399741) -- cycle;
\draw [line width=0.6pt][->] (0,0)-- (5,0);
\draw [line width=0.6pt][->] (0,0)-- (0,2);
\draw (-0.2,-0.05) node[anchor=north west] {$\text{O}$};
\draw (5,0.2) node[anchor=north west] {$x$};
\draw (-0.2,2.3) node[anchor=north west] {$y$};
\draw (1.6,0.5) node[anchor=north west] {$\text{M}$};
\draw (1.6,0) node[anchor=north west] {$\text{C}$};
\draw (0.55,0.26) node[anchor=north west] {$\alpha$};
\draw [line width=1pt] (1.80096,0.98109)-- (1.321997433360754,0);
\draw [line width=1pt] (1.321997433360754,0)-- (1.80096,-0.98109);
\draw [line width=1pt] (1.80096,-0.98109)-- (2.2799225666392458,0);
\draw [line width=1pt] (2.2799225666392458,0)-- (1.80096,0.98109);
\draw [line width=1pt,color=qqqqff] (0,0)-- (1.5057751530635666,0.37644378826589164);
\draw [line width=1pt,dash pattern=on 1pt off 1pt,color=qqqqff] (1.5057751530635666,0.37644378826589164)-- (2.0319285694405087,0.5079821423601272);
\draw [line width=1pt,color=qqqqff] (2.0319285694405087,0.5079821423601272)-- (4,1);
\draw [line width=1pt,color=qqqqff] (0,0)-- (1.5057751530635666,-0.37644378826589164);
\draw [line width=1pt,dash pattern=on 1pt off 1pt,color=qqqqff] (1.5057751530635666,-0.37644378826589164)-- (2.031928569440508,-0.507982142360127);
\draw [line width=1pt,color=qqqqff] (2.031928569440508,-0.507982142360127)-- (4,-1);
\draw [line width=0.6pt,dash pattern=on 1pt off 1pt] (4,-1)-- (4,1);
\begin{scriptsize}
\draw [fill=black] (1.80096,0) circle (1pt);
\draw [fill=black] (1.6839767500258735,0.42099418750646833) circle (1pt);
\draw [fill=black] (1.6839767500258735,-0.42099418750646833) circle (1pt);
\end{scriptsize}
\end{tikzpicture}
    \caption{Geometrical configuration of a spheroid rolling without slipping on diverging inclined rails. (a) 3D view. (b) Top view. (c) Side view.}
    \label{fig:enter-label}
\end{figure}

Let us now write the equation of the double-cone with the center $\mathrm{C}\,(x_\text{C}, 0, z_{\text{C}})$:
\[(x_\mathrm{M}-x_\mathrm{C})^2+(z_\mathrm{M}-z_\mathrm{C})^2=\dfrac{R^2}{h^2}(y_\mathrm{M}-h)^2.\]
We can rewrite this as a quadratic equation in terms of $x_\mathrm{M}$:
\begin{multline}
    x_\mathrm{M}^2 \lt 1+\tan^2 \theta - \dfrac{R^2}{h^2} \tan^2 \dfrac{\alpha}{2}\rt\\-2x_\mathrm{M}\lt x_\mathrm{C}+z_\mathrm{C} \tan\theta-\dfrac{R^2}{h}\tan\dfrac{\alpha}{2}\rt\\ + x_\mathrm{C}^2+z_\mathrm{C}^2-R^2=0. \label{eq11}
\end{multline}
The discriminant of equation (\ref{eq11}) is $\Delta_1$. Since $\mathrm{M}$ is the contact point, we can set the discriminant $\Delta_1$ to zero, which leads to:
\begin{multline}
    \lt x_\mathrm{C}+z_\mathrm{C}\tan\theta-\dfrac{R^2}{h}\tan\dfrac{\alpha}{2} \rt^2\\=\lt 1+\tan^2 \theta-\dfrac{R^2}{h^2}\tan^2 \dfrac{\alpha}{2}\rt \lt x_\mathrm{C}^2 +z_\mathrm{C}^2-R^2\rt.
    \label{eq12}
\end{multline}
Rewrite equation (\ref{eq12}) into a quadratic equation with respect to $z_{\text{C}}$:
\begin{multline*}
    z_\mathrm{C}^2\lt 1-\dfrac{R^2}{h^2}\tan^2 \dfrac{\alpha}{2}\rt \\-2z_\mathrm{C}\tan\theta\lt x_\mathrm{C}-\dfrac{R^2}{h}\tan \dfrac{\alpha}{2}\rt \\+x_\mathrm{C}^2 \lt \tan^2\theta -
    \dfrac{R^2}{h^2}\tan^2 \dfrac{\alpha}{2}\rt\\-R^2 \lt 1+\tan^2\theta-2\dfrac{x_\mathrm{C}}{h}\tan\dfrac{\alpha}{2}\rt=0.
\end{multline*}
This equation gives us the relation between $x_{\mathrm{C}}$ and $z_{\mathrm{C}}$:
\begin{equation}
    z_\mathrm{C}=\dfrac{\tan\theta \lt x_\mathrm{C}-\dfrac{R^2}{h}\tan\dfrac{\alpha}{2}\rt+\sqrt{\Delta}}{1-\dfrac{R^2}{h^2}\tan^2\dfrac{\alpha}{2}},
    \label{eq13}
\end{equation}
where 
\begin{multline*}
    \Delta=\tan^2\theta \lt x_\mathrm{C}-\dfrac{R^2}{h}\tan\dfrac{\alpha}{2}\rt^2\\+ R^2\lt 1-\dfrac{R^2}{h^2}\tan^2\dfrac{\alpha}{2}\rt\lt 1+\tan^2\theta-2\dfrac{x_\mathrm{C}}{h}\tan\dfrac{\alpha}{2}\rt\\-x_\mathrm{C}^2\lt 1-\dfrac{R^2}{h^2}\tan^2\dfrac{\alpha}{2}\rt\lt \tan^2\theta-\dfrac{R^2}{h^2}\tan^2\dfrac{\alpha}{2} \rt.
\end{multline*}
Notice that in Eq.\ref{eq13}, the plus sign is taken as the contact point should be the lower point.\\
Simplifying $\Delta$ gives us an interesting result:
\begin{multline}
    \Delta=R^2\lt\dfrac{x_\mathrm{C}}{h}\tan\dfrac{\alpha}{2}-1\rt^2\times\\\lt 1+\tan^2\theta-\dfrac{R^2}{h^2}\tan^2\dfrac{\alpha}{2}\rt.
    \label{eq14}
\end{multline}

Some straightforward geometric considerations yield an upper bound for \( x \), beyond which the double-cone would fall: $x_\mathrm{C}\leq h/\tan\frac{\alpha}{2}$, so:
\begin{equation*}
    \dfrac{x_\mathrm{C}}{h}\tan\dfrac{\alpha}{2}-1<0.
\end{equation*}
From Eq.\ref{eq13}, we can write $z_\mathrm{C}$ as a function of the position of the double-cone $x_\mathrm{C}$:
\begin{multline*}
    z_\mathrm{C}=\dfrac{\tan\theta\lt \dfrac{x_\mathrm{C}}{h}-\dfrac{R^2}{h^2}\tan\dfrac{\alpha}{2}\rt}{1-\dfrac{R^2}{h^2}\tan^2\dfrac{\alpha}{2}}\\+\dfrac{R}{h}\dfrac{\lt 1- \dfrac{x_\mathrm{C}}{h}\tan\dfrac{\alpha}{2} \rt\displaystyle\sqrt{1+\tan^2\theta-\dfrac{R^2}{h^2}\tan^2\dfrac{\alpha}{2}}}{1-\dfrac{R^2}{h^2}\tan^2\dfrac{\alpha}{2}}.
\end{multline*}
We can write $z_\mathrm{C}=A-Bx_\mathrm{C}$, with
\begin{equation}
    A=R\dfrac{\displaystyle\sqrt{1+\tan^2\theta-\dfrac{R^2}{h^2}\tan^2\dfrac{\alpha}{2}}-\dfrac{R}{h}\tan\theta\tan\dfrac{\alpha}{2}}{1-\dfrac{R^2}{h^2}\tan^2\dfrac{\alpha}{2}},
\end{equation}
\begin{equation}
    B=\dfrac{\dfrac{R}{h}\tan\dfrac{\alpha}{2}\displaystyle\sqrt{1+\tan^2\theta-\dfrac{R^2}{h^2}\tan^2\dfrac{\alpha}{2}}-\tan\theta}{1-\dfrac{R^2}{h^2}\tan^2\dfrac{\alpha}{2}}.
\end{equation}
From the results above, there exists a critical angle $\theta_0$ such that $z_\mathrm{C}$ remains constant $(B=0)$, and it can be calculated as follows:
\begin{equation}
    \tan\theta_0=\dfrac{R}{h}\tan\dfrac{\alpha}{2}.
    \label{eq17}
\end{equation}
The double-cone will ascend if $\theta<\theta_0$, as the potential energy is lower when $x_\mathrm{C}$ is larger.\\
Let $R/h=\tan\psi$, rewrite Eq.~\eqref{eq17} in $\psi, \phi$:
\begin{equation}
    \tan\theta_0=\tan\psi\tan\dfrac{\alpha}{2}.
    \label{eq18}
\end{equation}
Now we would want to find $\alpha, \theta$ in terms of the diverging angle $2\phi$, where $2\phi$ is the angle between the rails and it can be calculated as follows:\\
By taking the dot product between the two rails, we have:
\begin{equation}
    \cos2\phi=\dfrac{1-\tan^2\dfrac{\alpha}{2}+\tan^2\theta}{1+\tan^2\dfrac{\alpha}{2}+\tan^2\theta},
\end{equation}
so the angle $\phi$ can be calculated:
\begin{equation}
    \phi=\dfrac{1}{2}\cos^{-1}\lt\dfrac{1-\tan^2\dfrac{\alpha}{2}+\tan^2\theta}{1+\tan^2\dfrac{\alpha}{2}+\tan^2\theta}\rt.
\end{equation}
Applying the trigonometry \[\tan\lt\dfrac{1}{2}\cos^{-1}x\rt=\displaystyle\sqrt{\dfrac{{1-x}}{1+x}},\] we have:
\begin{equation}
    \tan\phi=\displaystyle\sqrt{\dfrac{\tan^2\dfrac{\alpha}{2}}{1+\tan^2\theta}}=\tan\dfrac{\alpha}{2}\cos\theta.
    \label{parameter}
\end{equation}
Combining Eq.~\ref{eq18} with Eq.~\ref{parameter}, we have:
\begin{equation}
    \dfrac{1}{\displaystyle\sqrt{1+\tan^2\theta_0}}=\cos\theta_0=\dfrac{\tan\phi}{\tan\dfrac{\alpha}{2}}=\dfrac{\tan\phi\tan\psi}{\tan\theta_0}.
\end{equation}
From the equation above, $\theta_0$ can be calculated in terms of $\psi$ and $\phi$ as follows:
\begin{equation}
    \tan\theta_0=\dfrac{\tan\phi\tan\psi}{\displaystyle\sqrt{1-\tan^2\phi\tan^2\psi}}.
\end{equation}
This result is consistent with the findings reported by Gandhi and Efthimiou~\cite{gandhi2005}.
\subsection{Equilibrium of the spheroid}
Now, we will apply the method above for the spheroid rolling upwards.\\
We consider a homogeneous spheroid of mass \( m \), with semi-minor axis \( a \) and semi-major axis \( b \), rolling without slipping on a pair of inclined rails, which diverge with an opening angle \( \alpha \) and are tilted at an inclination angle \( \theta \) relative to the horizontal plane. We assume that the major axis of the spheroid remains parallel to the \( y \)-axis at all times during the motion. Let the center of mass of the spheroid be denoted by $C=(x_{\mathrm{C}}, 0, z_{\mathrm{C}})$, and the contact point on the rail be $M = (x_M, y_M, z_M)$. Since there are two possible contact points symmetric with respect to the \( xz \)-plane, we choose the one with \( y_M > 0 \). Since M lies on the rail, its coordinates must satisfy the geometry of the rails. From Fig.~\ref{fig:geometry}, we obtain the following constraints on \( x_M \), \( y_M \), and \( z_M \):
\begin{align}
\label{eq01}
    y_M &=  x_M \tan \dfrac{\alpha}{2}, \\
    \label{eq02}
    z_M &= x_M \tan \theta.
\end{align}

\definecolor{xfqqff}{rgb}{0.498, 0, 1}
\definecolor{qqqqff}{rgb}{0, 0, 1}

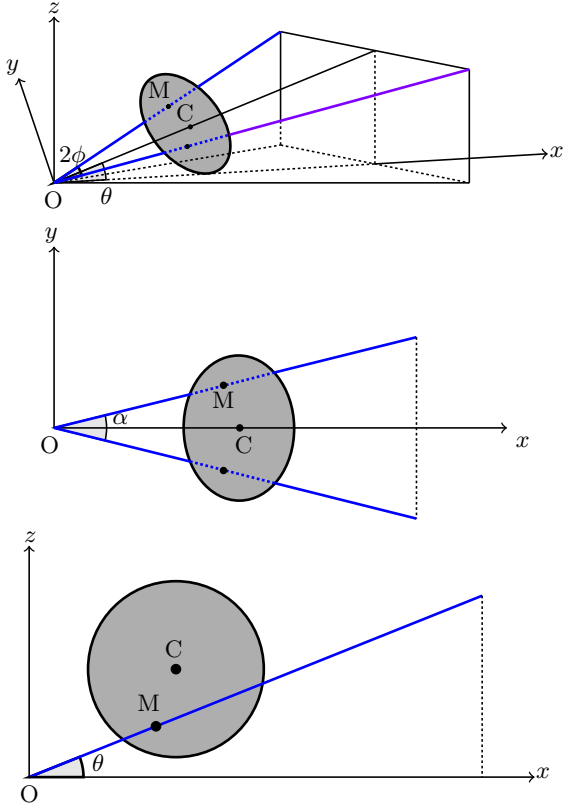
\begin{figure}[htbp]
    \centering

    % Hình (a) - 3D View
    \begin{minipage}[b]{0.5\textwidth}
        \centering
        \begin{tikzpicture}[scale=0.5]
            % --- TikZ picture 1 (3D View) ---
            \draw [shift={(-6,0)},line width=0.6pt,fill=black,fill opacity=0.1] (0,0) -- (3.37:1.38) arc (3.37:22.38:1.38) -- cycle;
            \draw [shift={(-6,0)},line width=1pt,fill=black,fill opacity=0.1] (0,0) -- (15.26:0.76) arc (15.26:33.69:0.76) -- cycle;
            \draw [line width=0.6pt] (-6,0)-- (5,0);
            \draw [line width=0.6pt] (5,0)-- (5,3);
            \draw [line width=0.6pt] (0,4)-- (5,3);
            \draw [line width=0.6pt] (-6,0)-- (2.5,3.5);
            \draw [line width=0.6pt,dash pattern=on 1pt off 1pt] (0,1)-- (5,0);
            \draw [line width=0.6pt,dash pattern=on 1pt off 1pt] (-6,0)-- (0,1);
            \draw [line width=0.6pt,dash pattern=on 1pt off 1pt] (-6,0)-- (2.5,0.5);
            \draw [line width=0.6pt,dash pattern=on 1pt off 1pt] (2.5,0.5)-- (2.5,3.5);
            \draw [line width=0.6pt] (0,4)-- (0,1.636);
            \draw [line width=0.6pt,dash pattern=on 1pt off 1pt] (0,1.636)-- (0,1);
            \draw [->, line width=0.6pt] (2.5,0.5)-- (7.08,0.77);
            \draw [->, line width=0.6pt] (-6,0)-- (-6.94,2.77);
            \draw [<- , line width=0.6pt] (-6,4.4)-- (-6,0);
            \draw [rotate around={-51.34:(-2.52,1.56)},line width=1pt,fill=black,fill opacity=0.32] (-2.52,1.56) ellipse (1.51cm and 0.93cm);
            \draw [line width=1pt,color=qqqqff] (-6,0)-- (-3.61,1.60);
            \draw [line width=1pt,dash pattern=on 1pt off 1pt,color=qqqqff] (-3.61,1.60)-- (-2.19,2.54);
            \draw [line width=1pt,color=qqqqff] (-2.19,2.54)-- (0,4);
            \draw [line width=1pt,color=qqqqff] (-6,0)-- (-3.09,0.79);
            \draw [line width=1pt,dash pattern=on 1pt off 1pt,color=qqqqff] (-3.09,0.79)-- (-1.36,1.27);
            \draw [line width=1pt,color=xfqqff] (-1.36,1.27)-- (5,3);
            \draw (-6.1,1.23) node[anchor=north west] {$2\phi$};
            \draw (6.9,1.2) node[anchor=north west] {$x$};
            \draw (-6.5,0) node[anchor=north west] {$\mathrm{O}$};
            \draw (-7.5,3.5) node[anchor=north west] {$y$};
            \draw (-6.4,5) node[anchor=north west] {$z$};
            \draw (-5,0.1) node[anchor=north west] {$\theta$};
            \draw (-3,2.3) node[anchor=north west] {$\mathrm{C}$};
            \draw (-3.75,2.9) node[anchor=north west] {$\mathrm{M}$};
            \begin{scriptsize}
                \draw [fill=black] (-2.39,1.48) circle (1.5pt);
                \draw [fill=black] (-2.97,2.02) circle (1.5pt);
                \draw [fill=black] (-2.47,0.96) circle (1.5pt);
            \end{scriptsize}
        \end{tikzpicture}
    \end{minipage}
    \hfill
    % Hình (b) - Top View
    \begin{minipage}[b]{0.5\textwidth}
        \centering
        \begin{tikzpicture}[scale=1.2]
            % --- TikZ picture 2 (Top View) ---
            \draw [shift={(0,0)},line width=0.6pt,fill=black,fill opacity=0.1] (0,0) -- (-14.04:0.58) arc (-14.04:14.04:0.58) -- cycle;
            \draw [->, line width=0.6pt] (0,0)-- (5,0);
            \draw [->, line width=0.6pt] (0,0)-- (0,2);
            \draw [rotate around={-89.22:(2.04,-0.0017)},line width=1pt,fill=black,fill opacity=0.32] (2.04,-0.0017) ellipse (0.80cm and 0.61cm);
            \draw [line width=1pt,color=qqqqff] (0,0)-- (1.50,0.37);
            \draw [line width=1pt,dash pattern=on 1pt off 1pt,color=qqqqff] (1.50,0.37)-- (2.43,0.61);
            \draw [line width=1pt,color=qqqqff] (2.43,0.61)-- (4,1);
            \draw [line width=1pt,color=qqqqff] (0,0)-- (1.50,-0.37);
            \draw [line width=1pt,dash pattern=on 1pt off 1pt,color=qqqqff] (1.50,-0.37)-- (2.44,-0.61);
            \draw [line width=1pt,color=qqqqff] (2.44,-0.61)-- (4,-1);
            \draw [line width=0.6pt,dash pattern=on 1pt off 1pt] (4,1)-- (4,-1);
            \draw (-0.25,0) node[anchor=north west] {$\mathrm{O}$};
            \draw (5,0) node[anchor=north west] {$x$};
            \draw (-0.2,2.3) node[anchor=north west] {$y$};
            \draw (1.65,0.5) node[anchor=north west] {$\mathrm{M}$};
            \draw (1.9,0) node[anchor=north west] {$\mathrm{C}$};
            \draw (0.55,0.25) node[anchor=north west] {$\alpha$};
            \begin{scriptsize}
                \draw [fill=black] (1.87,0.47) circle (1pt);
                \draw [fill=black] (1.87,-0.47) circle (1pt);
                \draw [fill=black] (2.05,0) circle (1pt);
            \end{scriptsize}
        \end{tikzpicture}
    \end{minipage}
    \hfill
    % Hình (c) - Side View
    \begin{minipage}[b]{0.5\textwidth}
        \centering
        \begin{tikzpicture}[scale=1.2]
            % --- TikZ picture 3 (Side View) ---
            \draw [line width=1pt,fill=black,fill opacity=0.32] (1.62,1.19) circle (0.97cm);
            \draw [shift={(0,0)},line width=1pt,fill=black,fill opacity=0.1] (0,0) -- (0:0.60) arc (0:21.80:0.60) -- cycle;
            \draw [line width=1pt,color=qqqqff] (0,0)-- (5,2);
            \draw [->, line width=0.6pt] (0,0)-- (5.58,0);
            \draw [->, line width=0.6pt] (0,0)-- (0,2.55);
            \draw [line width=0.6pt,dash pattern=on 1pt off 1pt] (5,2)-- (5,0);
            \draw (-0.2,0) node[anchor=north west] {$\mathrm{O}$};
            \draw (5.5,0.2) node[anchor=north west] {$x$};
            \draw (-0.15,2.8) node[anchor=north west] {$z$};
            \draw (0.6,0.35) node[anchor=north west] {$\theta$};
            \draw (1.4,1.6) node[anchor=north west] {$\mathrm{C}$};
            \draw (1.1,1) node[anchor=north west] {$\mathrm{M}$};
            \begin{scriptsize}
                \draw [fill=black] (1.62,1.19) circle (1.5pt);
                \draw [fill=black] (1.40,0.56) circle (1.5pt);
            \end{scriptsize}
        \end{tikzpicture}
      
    \end{minipage}

    \caption{Geometrical configuration of a spheroid rolling without slipping on diverging inclined rails. (a) 3D view. (b) Top view. (c) Side view.}
    \label{fig:geometry}
\end{figure}

Since the point M also lies on the surface of the spheroid, its coordinates \( (x_M, y_M, z_M) \) must satisfy the equation of the spheroid centered at \( C = (x_{\mathrm{C}}, 0, z_{\mathrm{C}}) \). Therefore, we have:
\begin{equation}
\label{eq03}
    \lt\dfrac{x_M - x_{\text{C}}}{a}\rt^2+\lt\dfrac{y_M}{b}\rt^2+\lt\dfrac{z_M - z_\text{C}}{a}\rt^2=1.
\end{equation}
Substituting Eqs.(\ref{eq01}) and (\ref{eq02}) into Eq.~(\ref{eq03}), we obtain the following quadratic equation in \( x_M \):

\begin{equation}
\label{eq1}
\splitfrac{x_M^2 \lt \dfrac{1}{a^2}+\dfrac{\tan^2 \dfrac{\alpha}{2}}{b^2}+\dfrac{\tan^2 \theta}{a^2} \rt}{-2x_M \dfrac{x_\text{C}+z_\text{C}\tan\theta}{a^2}+ \lt \dfrac{x_{\text{C}}^2 + z_{\text{C}}^2 -a^2}{a^2} \rt =0.}
\end{equation}

Since M is the unique contact point between the spheroid and the rail, Eq.~(\ref{eq1}) must have a double root at \(x_{M}\). Thus, its discriminant must be zero, leading to the following condition:

\begin{equation}
    \label{eq2}
\splitfrac{(x_{\text{C}} +z_{\text{C}} \tan \theta)^2=}{\lt 1+\tan^2 \theta + \dfrac{a^2}{b^2} \tan^2 \dfrac{\alpha}{2} \rt (x_{\text{C}}^2 +z_{\text{C}}^2- a^2).}
\end{equation}
Eq.~(\ref{eq2}) can be rearranged into a quadratic form with respect to \( z_{\text{C}} \).
   \begin{align}
    &z_{\text{C}}^2 \left( \dfrac{a^2}{b^2}\tan^2 \dfrac{\alpha}{2} +1 \right) 
    - 2z_\text{C} x_{\text{C}} \tan \theta \notag \\
    &\quad + x_{\text{C}}^2 \left( \tan^2 \theta + \dfrac{a^2}{b^2} \tan^2 \dfrac{\alpha}{2} \right) \notag \\
    \label{eq06}
    &-a^2 \left( \dfrac{a^2}{b^2}\tan^2 \dfrac{\alpha}{2} +1 +\tan^2 \theta \right)=0.
\end{align}
Solving Eq. (\ref{eq06}) gives us the relation between $z_{\text{C}}$ and $x_{\text{C}}$:
\begin{equation}
    \label{eq4}
    z_{\text{C}} = \dfrac{x_{\text{C}} \tan \theta + \sqrt{\Delta}}{1 + \dfrac{a^2}{b^2} \tan^2 \dfrac{\alpha}{2}},
\end{equation}
where 
\begin{equation}
    \splitfrac{\Delta = \lt 1+ \tan^2 \theta + \dfrac{a^2}{b^2} \tan^2 \dfrac{\alpha}{2} \rt \times}{\left[ a^2 \lt 1 + \dfrac{a^2}{b^2} \tan^2 \dfrac{\alpha}{2} \rt -x_{\text{C}}^2 \lt \dfrac{a^2}{b^2} \tan^2 \dfrac{\alpha}{2} \rt \right ].}
\end{equation}

\begin{figure}[htbp]
    \centering
    \includegraphics[width=0.48\textwidth]{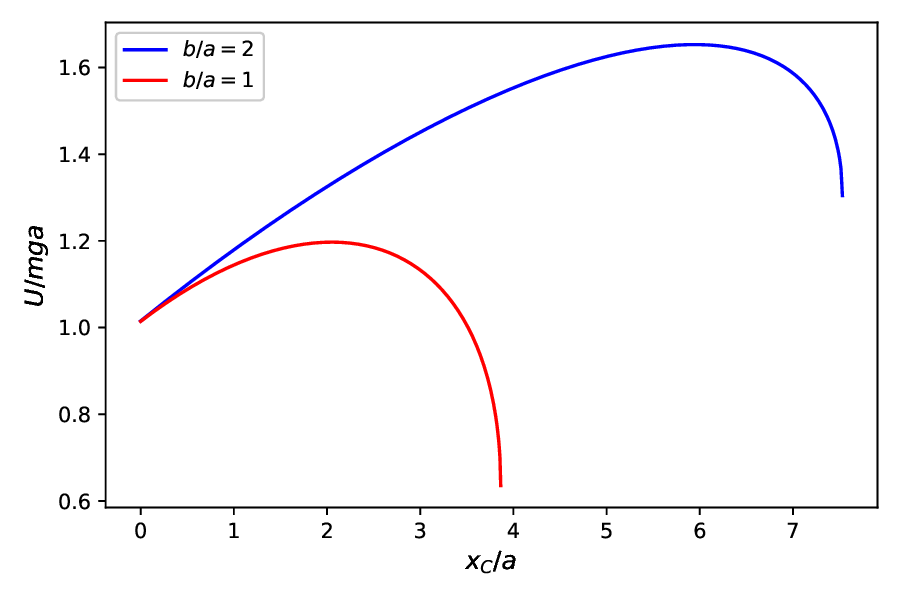}
    \caption{The potential energy of the ellipsoid versus its position}
    \label{fig4}
\end{figure}

The expression for \( z_{\text{C}} \) obtained in Eq.~(\ref{eq4}) allows us to determine the potential energy of the spheroid as a function of \( x_{\text{C}} \), taking the origin O as the reference point:
\begin{equation}
    U(x_{\text{C}})=mg z_{\text{C}}=mg \dfrac{x_{\text{C}} \tan \theta + \sqrt{\Delta}}{1 + \dfrac{a^2}{b^2} \tan^2 \dfrac{\alpha}{2}},
\end{equation}
where \(g\) denotes the gravitational acceleration. The FIG.~\ref{fig4} shows the potential energy \(U(x_{\text{C}})\) for various values of the ratio \(b/a\), where the parameters are set to \( \alpha = 30^\circ \) and \( \theta = 10^\circ \). From the graph of \( U(x_{\text{C}}) \), we observe the existence of an unstable equilibrium point \( x_{\text{eq}} \). If the spheroid is released from rest at an initial position \( x_0 > x_{\text{eq}} \), it will continue to roll upward along the rails,  \( x_{\text{C}} \) keeps increasing. In contrast, if it is released from rest at an initial position \( x_0 < x_{\text{eq}} \), the spheroid will roll downward, corresponding to a decreasing \( x_{\text{C}} \).

By solving the equation \( \displaystyle \dfrac{\mathrm{d}U}{\mathrm{d}x_{\text{C}}} = 0 \), we obtain the unstable equilibrium position \( x_{\text{eq}} \):
\begin{equation}
    x_{\text{eq}}=\dfrac{a \tan \theta}{\dfrac{a}{b}\tan \dfrac{\alpha}{2}{\sqrt{\dfrac{a^2}{b^2}\tan^2 \dfrac{\alpha}{2}+\tan^2 \theta }}}.
    \label{x_0}
\end{equation}
\begin{figure}[htbp]
    \centering
    \includegraphics[width=0.48\textwidth]{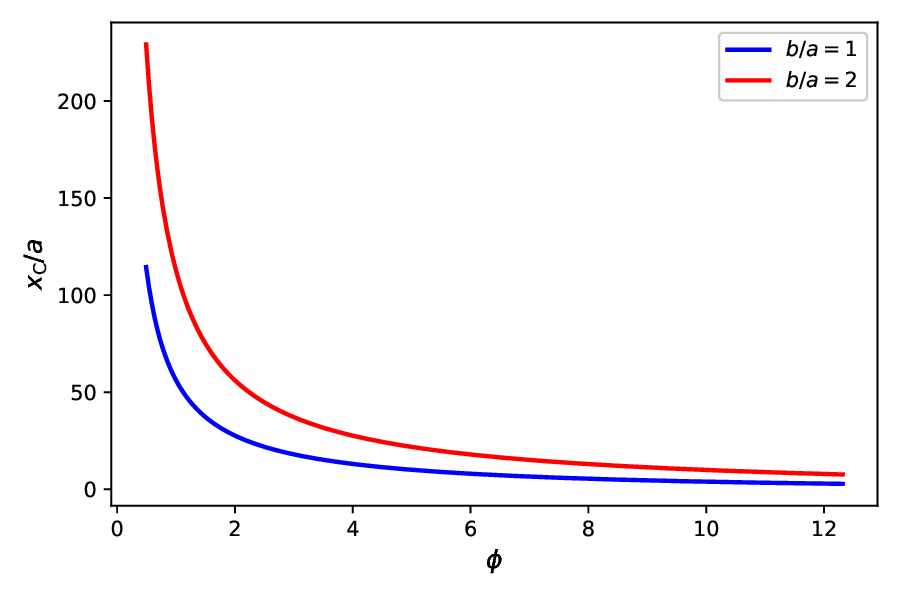}
    \caption{The equilibrium point of the ellipsoid versus the diverged angle $\phi$ of the rails, with the value of $\theta$ is fixed to $\theta=1^\circ$.}
    \label{xeqalpha}
\end{figure}
\begin{figure}[htbp]
    \centering
    \includegraphics[width=0.48\textwidth]{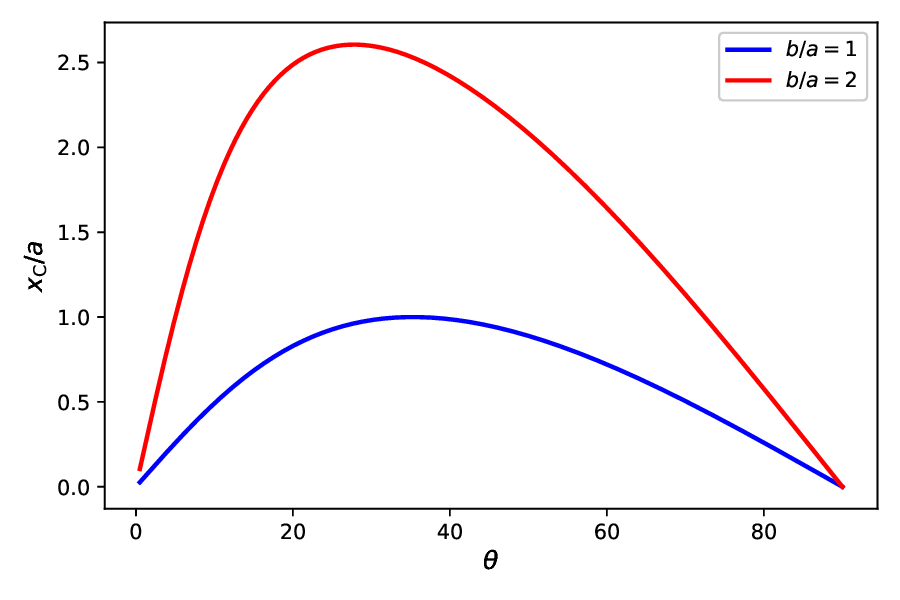}
    \caption{The equilibrium point of the ellipsoid versus the inclined angle $\theta$ of the rails, with the value of $\phi$ is fixed to $\phi=30^\circ$.}
    \label{xeqtheta}
\end{figure}\\
FIG.~\ref{xeqalpha} and FIG.~\ref{xeqtheta} show the dependencies of the equilibrium point on two parameters $\phi$ and $\theta$ of the rails.\\
We will now consider the point at which the spheroid falls down. Since $\Delta \geq 0$, we have: 
\begin{equation}
    {x_{\max}} = \sqrt {{b^2}{{\cot }^2}\frac{\alpha }{2} + {a^2}}.
\end{equation}
Therefore, at \begin{equation}
    x_\text{e}=b\sqrt{\dfrac{1}{\tan^2 \dfrac{\alpha}{2}}+\dfrac{a^2}{b^2}}=b\displaystyle\sqrt{\dfrac{\cos^2\theta}{\tan^2\phi}+\dfrac{a^2}{b^2}},
\end{equation} the spheroid will fall down.\\
\subsection{The dynamics of the rolling spheroid}
We have discussed the equilibrium state of the spheroid on the rails. We now proceed to analyze the motion of the spheroid as it rolls upward from the equilibrium position. Throughout this investigation, we consider the motion to be pure rolling. Under this condition, the spheroid rotates about the instantaneous point of contact \( M \) during its motion.\\
The spheroid is placed at the equilibrium point, thus the initial potential energy of the spheroid can be expressed as:
\begin{equation}
    U_0=U(x_\text{eq})=mg\dfrac{Bx_\text{eq}+\sqrt{\Delta_0}}{1+A^2},
\end{equation}
where $A=\dfrac{a}{b}\tan\dfrac{\alpha}{2},\,B=\tan\theta$.\\
We can calculate $x_\text{eq}$ and $\Delta_0$ in terms of $A$ and $B$ as follows:
\begin{align}
    x_\text{eq}&=a\dfrac{B}{A\sqrt{A^2+B^2}},\\
    \Delta_0&=\dfrac{A^2\lt1+A^2+B^2\rt^2}{A^2+B^2}.
\end{align}
Accordingly, the initial potential of the spheroid is
\begin{equation}
    U_0=mga\dfrac{\sqrt{A^2+B^2}}{A}.
\end{equation}
Consider the spheroid located at a position \( x_\mathrm{C} \), where \( x_\mathrm{C} \geq x_\text{eq} \). Let the angular velocity of the sphere vector be $\boldsymbol{\omega}$, the velocity of the center of mass $\mathrm{C}$: $v_\text{C}= \omega R_*$, where $R_*$ is the distance from $\mathrm{M}$ to the pivot, and the vector $\boldsymbol{R_*}$ points to $\mathrm{M}$. We will use $\hat{\boldsymbol{x}}, \hat{\boldsymbol{y}},\hat{\boldsymbol{z}}$ to represent the unit vectors of $\mathrm{O}x, \mathrm{O}y, \mathrm{O}z$, then $\boldsymbol{\omega}=\omega \hat{\boldsymbol{y}}$.\\
Since $\hat{\boldsymbol{y}}$ is the unit vector of the pivot, we have: 
\begin{equation}
    R_*=\dfrac{|\mathbf{M C}\times \hat{\boldsymbol{y}}|}{|\hat{\boldsymbol{y}}|}=|\mathbf{MC}\times \hat{\boldsymbol{y}}|,
    \label{r*}
\end{equation}
The vector $\mathrm{MC}$ is: \[\mathbf{MC}=(x_\mathrm{C}-x_\mathrm{M})\hat{\boldsymbol{x}}+(y_\mathrm{C}-y_\mathrm{M})\hat{\boldsymbol{y}}+(z_\mathrm{C}-z_\mathrm{M})\hat{\boldsymbol{z}}.\] Taking the cross product between $\mathrm{MC}$ and $\hat{\boldsymbol{y}}$, we get:
\begin{equation}
        \mathbf{MC}\times\hat{\boldsymbol{y}}=(x_\mathrm{C}-x_\mathrm{M})\hat{\boldsymbol{z}}-(z_\mathrm{C}-z_\mathrm{M})\hat{\boldsymbol{x}}.
\end{equation}
From Eq.~\eqref{r*}, we have:
\begin{equation}
    R_*=\sqrt{\lt x_\mathrm{C}-x_\mathrm{M}\rt^2+\lt z_\mathrm{C}-z_\mathrm{M}\rt^2}.
\end{equation}
It is easily seen from Eq.~(\ref{eq01}) and Eq.~(\ref{eq03}) that
\begin{equation}
    \lt x_\mathrm{C}-x_\mathrm{M}\rt^2+\lt z_\mathrm{C}-z_\mathrm{M}\rt^2=a^2-y_\mathrm{M}^2\dfrac{a^2}{b^2}=a^2-A^2 x_M^2.
\end{equation}
Solving the Eq.~\ref{eq1}, we have:
\begin{equation}
    x_\mathrm{M}=\dfrac{x_\mathrm{C}+Bz_\mathrm{C}}{1+A^2+B^2}.
\label{xm}
\end{equation}
Substituting $z_\mathrm{C}$ from Eq.~\ref{eq4} into Eq.~\ref{xm} yields an expression for $R_*$ of $x_\mathrm{C}$:
\begin{equation}
    R_*=\sqrt{a^2-A^2 \dfrac{\lt{x_C+B\dfrac{Bx_\mathrm{C}+\sqrt{\Delta}}{1+A^2}}\rt^2}{\lt 1+A^2+B^2 \rt^2}},
\end{equation}
which leads to the following relation:
\begin{equation}
    v_\mathrm{C}=\omega\sqrt{a^2-A^2 \dfrac{\lt{x_C+B\dfrac{Bx_\mathrm{C}+\sqrt{\Delta}}{1+A^2}}\rt^2}{\lt 1+A^2+B^2 \rt^2}}.
\end{equation}
The kinetic energy of the sphere, assuming a moment of inertia 
 $I=\dfrac{2}{5}ma^2$, is given by:
\begin{equation*}
    K=\dfrac{1}{2}mv_\mathrm{C}^2+\dfrac{1}{2}I\omega^2.
\end{equation*}
This can be expressed solely in terms of the velocity of the center of mass $v_\mathrm{C}$:
\begin{equation}
K=\dfrac{1}{10}mv_\mathrm{C}^2\dfrac{7a^2-5\lt Dx_\mathrm{C}+
E\sqrt{a^2(1+A^2)^2-x_\mathrm{C}^2}\rt^2}{a^2-\lt Dx_\mathrm{C}+E\sqrt{a^2(1+A^2)^2-x_\mathrm{C}^2}\rt^2},
\end{equation}
where $D=\dfrac{A}{1+A^2},\,E=\dfrac{ABD}{\sqrt{(1+A^2)(1+A^2+B^2)}}.$\\
Under pure-rolling at contact, the friction does no work and the mechanical energy is conserved. Therefore, we have
\(
U_0 = K + U,
\)
where \( U_0 \) is the initial potential energy at equilibrium. From this relation, the velocity of the center of mass \( v_\mathrm{C} \) can be expressed as a function of its position \( x_\mathrm{C} \):
\begin{equation}
    \splitfrac{v_\mathrm{C}^2=10g\lt a\dfrac{\sqrt{A^2+B^2}}{A}-\dfrac{x_\mathrm{C}B+\sqrt{\Delta}}{1+A^2}\rt\times}{\dfrac{a^2-\lt Dx_\mathrm{C}+E\sqrt{a^2(1+A^2)^2-x_\mathrm{C}^2}\rt^2}{7a^2-5\lt Dx_\mathrm{C}+
E\sqrt{a^2(1+A^2)^2-x_\mathrm{C}^2}\rt^2}.}
\label{vc}
\end{equation}
The center-of-mass velocity decomposes into two components,
\(v_{\mathrm{C}x}=\dot{x}_{\mathrm{C}}\) and \(v_{\mathrm{C}z}=-\dot{z}_{\mathrm{C}}\);
using Eq.~\eqref{eq4} gives

\begin{equation}
    v_{\mathrm{C}z}=v_{\mathrm{C}x}\lt \dfrac{ADx_\mathrm{C}\lt 1+A^2+B^2\rt}{\sqrt{\Delta}}-\dfrac{B}{1+A^2}\rt.
    \label{vxz}
\end{equation}
Combining this with the relation \(v_{\text{C}x}^2+v_{\text{C}z}^2=v_\text{C}^2,\) we can find both $v_{\mathrm{C}x}$ and $v_{\mathrm{C}z}$ as functions of $x_\mathrm{C}$.\\
\subsection{A special case: Sphere rolling upwards}
When $a=b=R$, the spheroid will become a sphere. In this case, we have a geometric approach to find $x_\text{e}$:
As the sphere descends, the line becomes tangent to its surface. Consequently, the relationship can be expressed as:
\(
x_\text{e} \sin \frac{\alpha}{2} = R.
\)
Solving for \( x_\text{e} \), we obtain:
\begin{equation}
    x_\text{e} = \dfrac{R}{\sin \dfrac{\alpha}{2}}=R\displaystyle\sqrt{1+\dfrac{\cos^2\theta}{\tan^2\phi}}.
\end{equation}
This result is consistent with the analytical expression for \( x_\text{e} \).

Consider the rolling process of the sphere when $\theta=0^\circ$. From Eq.~\ref{vc}, we have the velocity of $\mathrm{C}$ in terms of $x_\mathrm{C}$:
\begin{equation}
\label{eq5}
v_\text{C}=\sqrt{\frac{40R^2-10x_\mathrm{C}^2 \sin^2 \alpha}{28R^2-5x_\mathrm{C}^2 \sin^2 \alpha}\lt R-\sqrt{R^2-x_\mathrm{C}^2 \sin^2 \dfrac{\alpha}{2}} \rt g}\,.
\end{equation}
Moreover, from Eq.~\ref{vxz}, we get the relationship between the two components of the velocity of the sphere as follows:
\begin{equation}
    v_{\text{C}z}=v_{\text{C}x} \dfrac{x_\text{C}\sin^2 \dfrac{\alpha}{2}}{\sqrt{R^2-x_\text{C}^2 \sin^2 \dfrac{\alpha}{2}}}.
    \label{vxzs}
\end{equation}
Combining this with the relation \(v_{\text{C}x}^2+v_{\text{C}z}^2=v_\text{C}^2,\) we can deduce:
\begin{equation}
\label{eq6}
\splitfrac{v_{\text{C}}=}{\sqrt{\dfrac{40 \lt R^2-x_\text{C}^2\sin^2 \dfrac{\alpha}{2}\rt \lt R-\sqrt{R^2-x_\text{C}^2 \sin^2 \dfrac{\alpha}{2}}\rt}{28R^2-5x_\text{C}^2 \sin^2 \alpha}g}\,.}
\end{equation}
Since $\dot{x}_\text{C}=v_\text{C}$, taking the integral of (\ref{eq6}) from the beginning point S $(x_\text{S}=s)$ to $x_\text{C}$, we can express the moving time $t$ of the sphere in that process by the integral:
\begin{equation}
    \label{eq8}
{\int_{s}^{x_\mathrm{C}}{\frac{\mathrm{d}x_\mathrm{C}}{\displaystyle\sqrt{\frac{40 \lt R^2-x_\mathrm{C}^2 \sin^2 \dfrac{\alpha}{2} \rt \lt R-\sqrt{R^2-x_\mathrm{C}^2 \sin^2 \dfrac{\alpha}{2}}\rt}{g\lt28R^2-5x_\mathrm{C}^2 \sin^2 \alpha\rt}}}}}.
\end{equation}
Notice that the initial point must be chosen at S with $x_\mathrm{S} > 0$, otherwise, the sphere will stabilize at O.

\begin{figure}[htbp]
    \centering
    \includegraphics[width=0.48\textwidth]{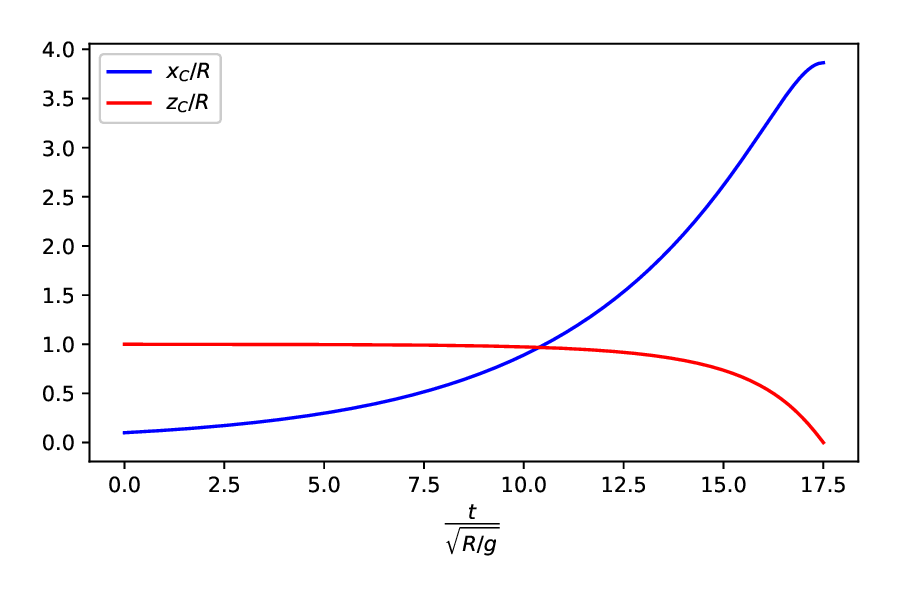}
    \caption{The position of the sphere versus time}
    \label{fig1}
\end{figure}

\begin{figure}[htbp]
    \centering
    \includegraphics[width=0.48\textwidth]{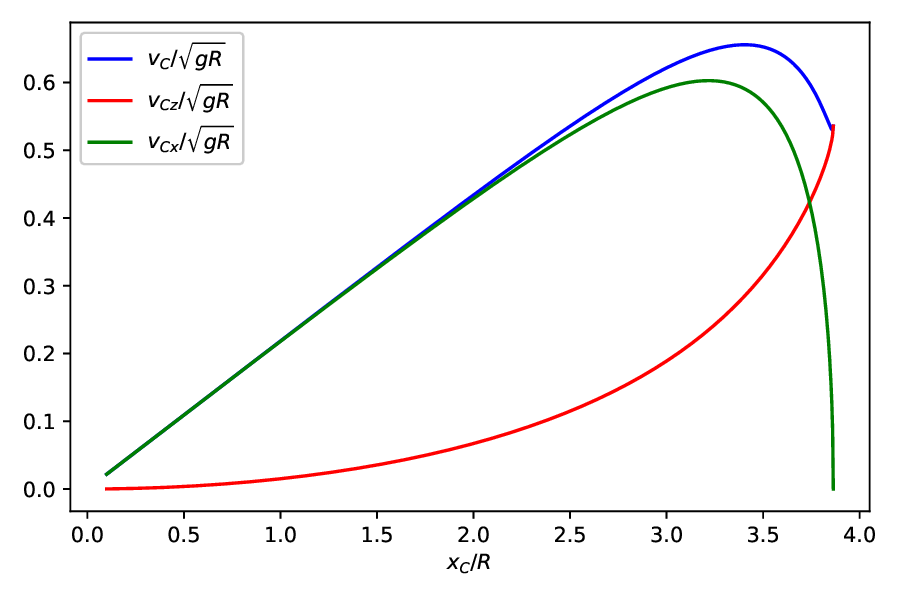}
    \caption{The velocities of the sphere versus its position}
    \label{fig2}
\end{figure}

\begin{figure}[htbp]
    \centering
    \includegraphics[width=0.48\textwidth]{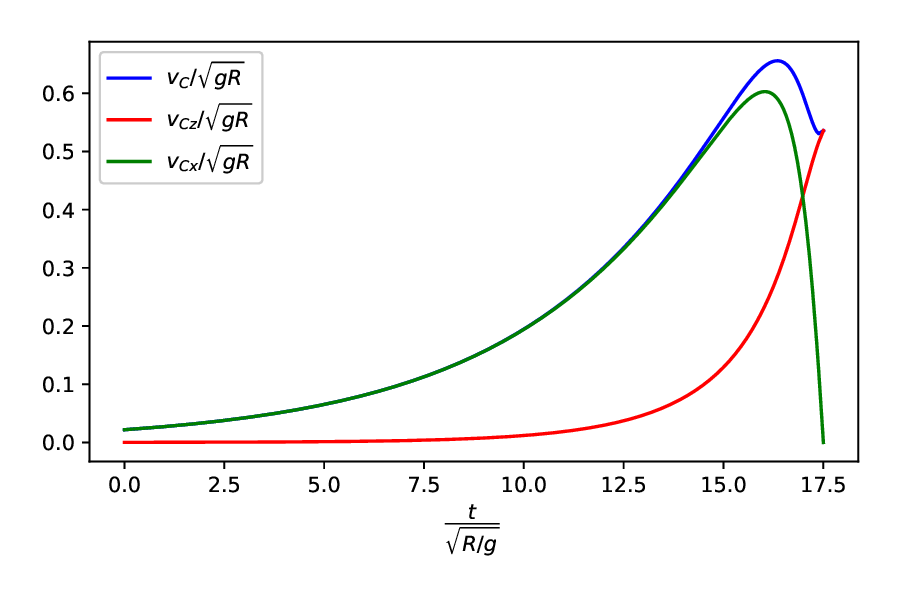}
    \caption{The velocities of the sphere versus time}
    \label{fig3}
\end{figure}

Moreover, from Eq. (\ref{eq5}) and Eq. (\ref{vxzs}), we can find $v_{\text{C}z}$ also in terms of $x_\mathrm{C}$:
\begin{equation}
    \label{eq9}
    \dfrac{v_{\text{C}z}}{\displaystyle\sqrt{g/R}}=x_\mathrm{C}\sin^2 \dfrac{\alpha}{2}\displaystyle\sqrt{\dfrac{40\lt R-\displaystyle\sqrt{R^2-x_\mathrm{C}^2 \sin^2 \dfrac{\alpha}{2}}\rt}{28R^2-5x_\mathrm{C}^2 \sin^2 \alpha}}\,.
\end{equation}
From equations (\ref{eq5}), (\ref{eq6}), (\ref{eq8}), and (\ref{eq9}), we obtain three corresponding plots, shown in FIG. \ref{fig1}, FIG. \ref{fig2}, and FIG. \ref{fig3}, respectively: $x_\text{C},z_\text{C}$ versus $t$, $v_\text{C}, v_{\text{C}x}, v_{\text{C}z}$ versus $x$ and $v_\text{C}, v_{\text{C}x}, v_{\text{C}z}$ versus $t$, with the numerical values: $s=0.1R$, $\phi=60^\circ$.

\section{Experiment}
\begin{figure}[htbp]
    \centering
    \includegraphics[width=0.3\textwidth]{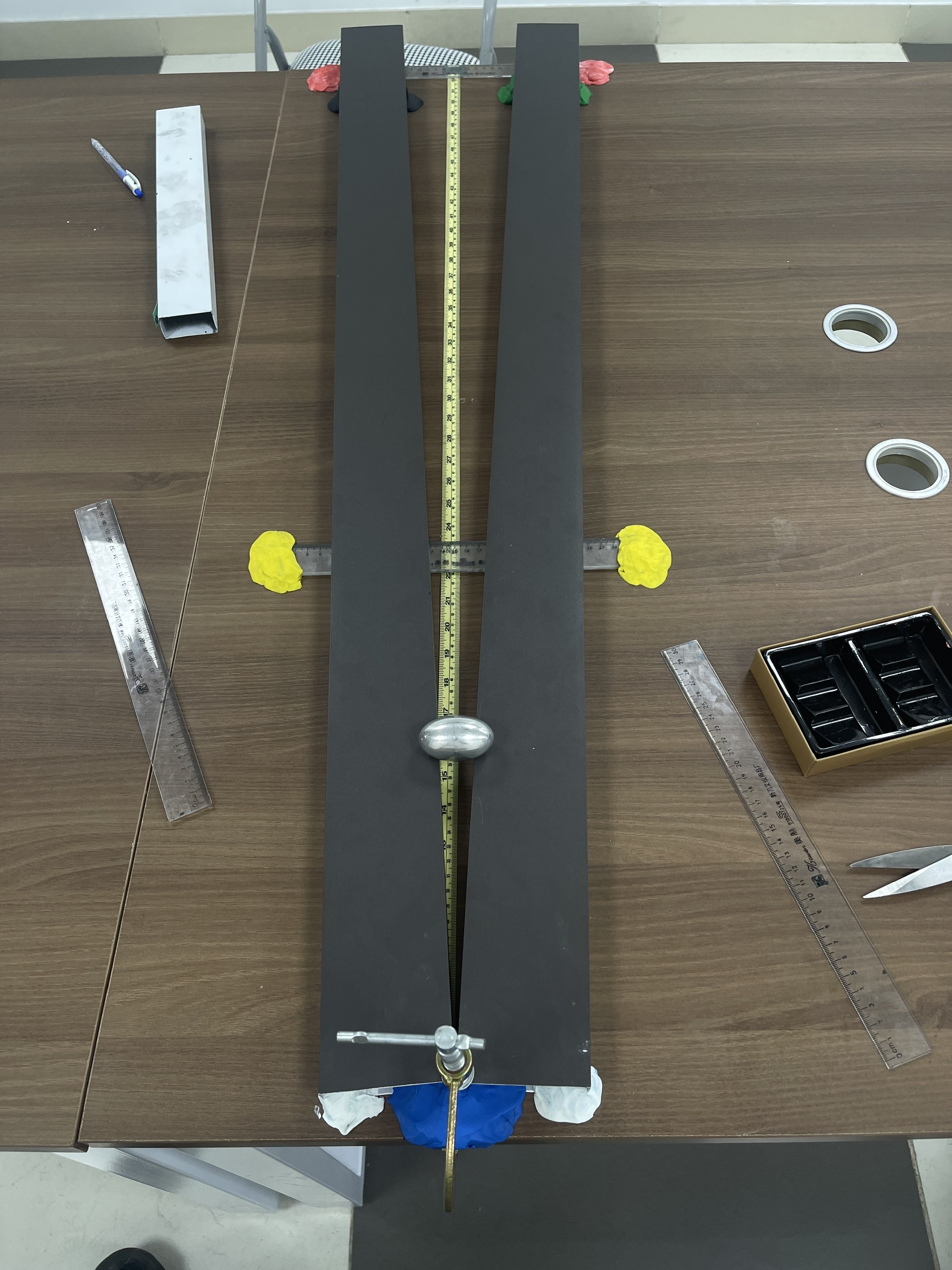}
    \caption{Experimental setup showing the three-bar configuration.}
    \label{fig_4}
\end{figure}

The experimental results obtained during the course of this investigation are presented in this section. These data have been systematically analyzed and compared with the theoretical predictions outlined in the preceding sections.\\
We consider the equilibrium point \( ax_0 \), as defined in Equation~\ref{x_0}. The experimental apparatus, shown in Figure~\ref{fig_4}, consists of three bars: two serve as parallel rails on which a spheroid rolls, and the third is used to adjust the inclination angle \(\theta\). The spheroid used in this experiment has dimensions of \(2 \times 1.25 \times 1.25\) inches.\\
Two independent experiments were conducted:
\begin{itemize}
    \item In Experiment 1, the divergence angle \(\alpha\) was held constant while the inclination angle \(\theta\) was varied.
    \item In Experiment 2, \(\theta\) was held constant while \(\alpha\) was varied.
\end{itemize}

\begin{figure}[H]
    \centering
    \includegraphics[width=0.4\textwidth]{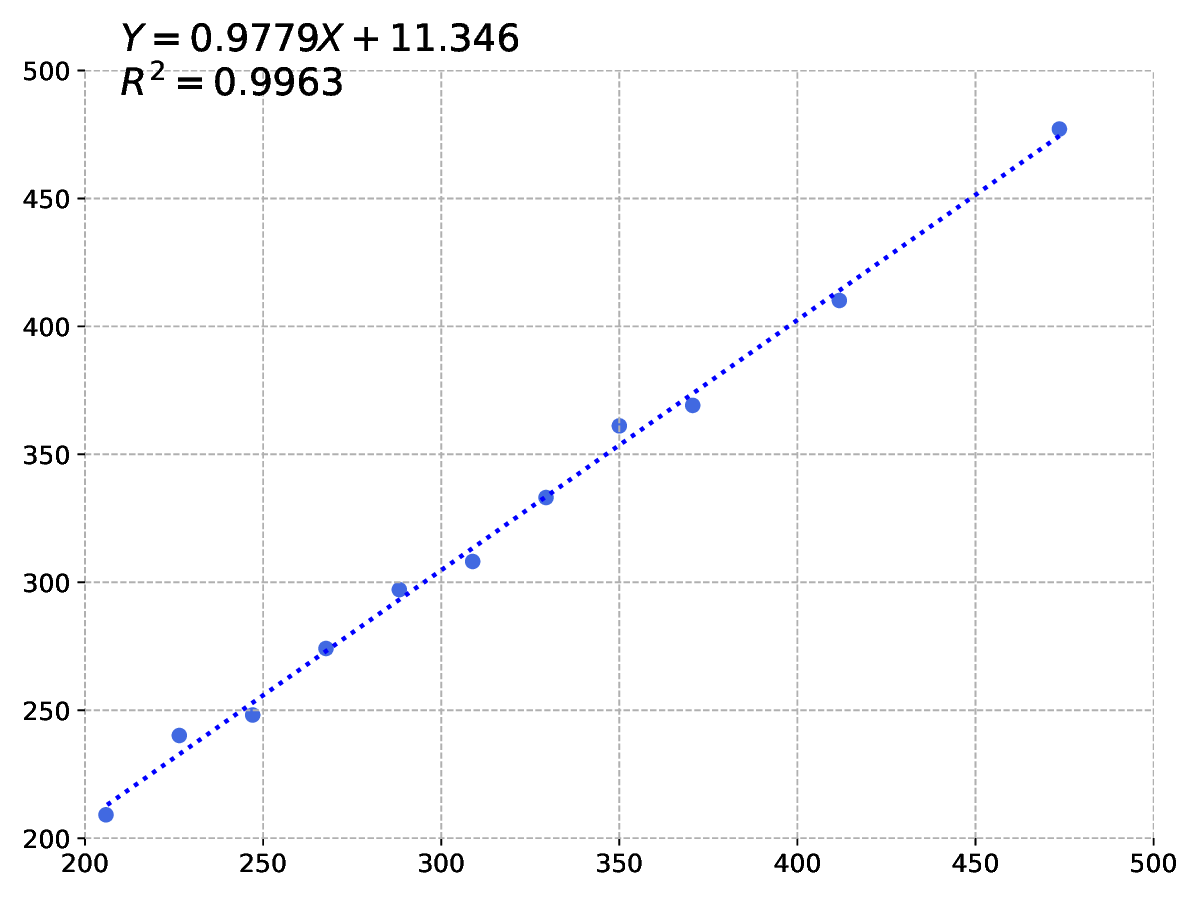}
    \caption{Experimental data with varying inclination angle $\theta$ and fixed divergence angle $\alpha$.}
    \label{fig_5}
\end{figure}

\begin{figure}[H]
    \centering
    \includegraphics[width=0.4\textwidth]{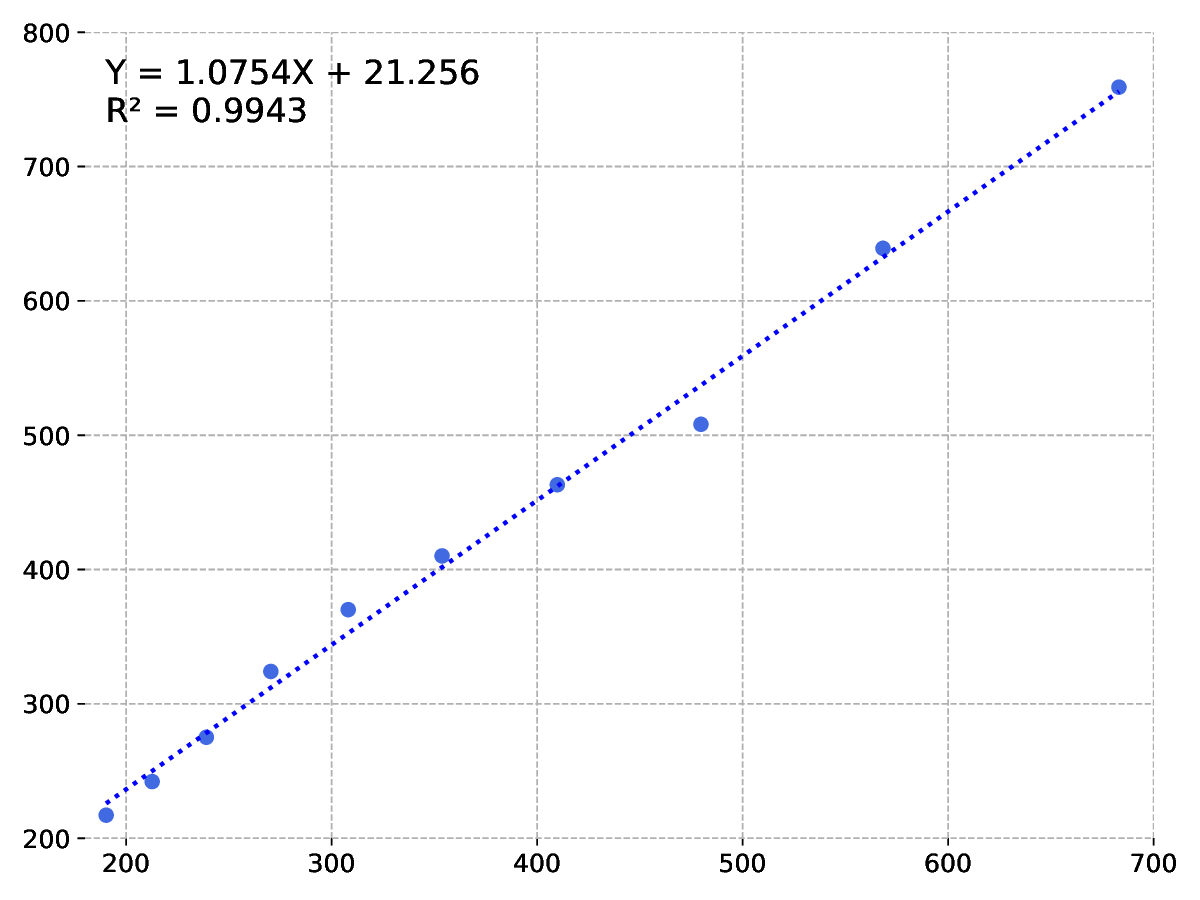}
    \caption{Experimental data with varying divergence angle $\alpha$ and fixed inclination angle $\theta$.}
    \label{fig_6}
\end{figure}

The experimental parameters are defined as follows:
\begin{itemize}
    \item length $L$ represents the position of the third bar, which determines the inclination angle $\theta$,
    \item $H$ denotes the height of the third bar, measured at the position corresponding to length $L$, $H=300\,\text{mm}$,
    \item $D$ denotes the length between the two intersection points where the rolling rails make contact with the third bar, which determines the angle $\alpha$.
    \item $x$ denotes the maximum equilibrium position of the system’s center of mass, which should exceed the theoretical value due to friction.
    \item $X$ denotes the theoritical value of the centre of mass,
    \item $Y=x$.
\end{itemize}
Based on these definitions, the relationship between $X$ and $Y$ is expected to be linear with a slope of approximately 1 and a positive $y$-intercept, attributable to unmodeled effects such as rolling friction and surface imperfections.\\
Figure~\ref{fig_5} shows the data from Experiment 1 with a fixed divergence $D = 45\,\mathrm{mm}$, where $\theta$ was varied.\\
Figure~\ref{fig_6} presents the results from Experiment 2, where the divergence angle $\alpha$ was varied while holding the inclination length $L = 1165\,\mathrm{mm}$ constant.


In both experiments, the data exhibit a strong linear correlation between the measured equilibrium position $Y$ and the theoretical value $X$. The observed slope is close to unity, while the consistent presence of a positive $y$-intercept supports the theoretical model when minor dissipative effects are considered. These findings confirm the validity of the geometric and energetic analysis presented earlier and demonstrate the robustness of the experimental setup.
\section*{Acknowledgements}
% I would like to express my gratitude towards my primary advisor Duy V. Nguyen for first suggesting the problem studied here and for incisive guidance during its entire development. 
We thank Phenikaa University for authorizing the experiments and providing access to laboratory space and equipment; the staff’s help with logistics, safety, and setup materially improved the quality and reliability of the measurements. Finally, Khanh P. M. Hoang wants to thank his family for steady encouragement throughout this work. 

\bibliographystyle{apsrev4-2}
\bibliography{ref}

\end{document}